\documentclass[aps,preprintnumbers,eqsecnum,amsmath,amssymb,showpacs,nofootinbib]{revtex4}
\usepackage{graphicx}
\usepackage{dcolumn}
\usepackage{bm}
\usepackage{epsfig}

\begin{document}
\title{Systematic uncertainties of hadron parameters obtained with QCD sum rules}  
\author{Wolfgang Lucha$^{a}$, Dmitri Melikhov$^{a,b}$ and Silvano Simula$^{c}$}
\affiliation{
$^a$ Institute for High Energy Physics, Austrian Academy of Sciences, Nikolsdorfergasse 18, A-1050, Vienna, Austria\\
$^b$ Nuclear Physics Institute, Moscow State University, 119992, Moscow, Russia\\
$^c$ INFN, Sezione di Roma III, Via della Vasca Navale 84, I-00146, Roma, Italy}
\date{\today}
\begin{abstract}
We study the uncertainties of the determination of the ground-state parameters from 
Shifman-Vainshtein-Zakharov 
(SVZ) sum rules, making use of the harmonic-oscillator potential model as an example. 
In this case, one knows the exact solution for the polarization operator 
$\Pi(\mu)$, which allows one to obtain both the OPE to any order and the 
spectrum of states.  
We start with the OPE for $\Pi(\mu)$ and analyze the extraction of 
the square of the ground-state wave function, $R\propto|\Psi_0(\vec r=0)|^2$, 
from an SVZ sum rule, setting the mass of the ground state $E_0$ equal to its known 
value and treating the effective continuum threshold as a fit parameter. 
We show that in a limited ``fiducial'' range of the Borel parameter
there exists a solution for the effective threshold which {\it precisely} 
reproduces the exact $\Pi(\mu)$ for any value of $R$ within the range  
$0.7 \le R/R_0 \le 1.15$ ($R_0$ is the known exact value). 
Thus, the value of $R$ extracted from the sum rule is determined to a great 
extent by the contribution of the hadron continuum. Our main finding is that
in the cases where the hadron continuum is not known and is modeled by an effective 
continuum threshold, the systematic uncertainties of the sum-rule procedure cannot be controlled.  
\end{abstract}
\maketitle

\section{Introduction}
A QCD sum-rule calculation of hadron parameters \cite{svz} 
involves two steps: (i) one calculates the operator product expansion (OPE) series 
for a relevant correlator, and (ii) one extracts the parameters  
of the ground state by a numerical procedure. 
Each of these steps leads to uncertainties in the final result. 

The first step lies fully within QCD and allows a rigorous treatment of the uncertainties:  
the correlator in QCD is not known precisely (because of 
uncertainties in quark masses, condensates, $\alpha_s$, radiative corrections, etc), but the 
corresponding errors in the correlator may be systematically controlled (at least in principle).

The second step lies beyond QCD and is more cumbersome: even if several terms 
of the OPE for the correlator were known precisely, the hadronic parameters might be extracted 
by a sum rule only within some error, which may be treated as a systematic error of the method. 
It is useful to recall that a successful extraction of the hadronic parameters by a sum 
rule is not guaranteed: as noticed already in the classical papers \cite{svz,nsvz}, 
the method may work in some cases and fail in others; moreover, error estimates 
(in the mathematical sense) for the numbers obtained by sum rules may not be easily provided --- 
e.g., according to \cite{svz}, any value obtained by varying the parameters in the sum-rule 
stability region has equal probability. 
However, for many applications of sum rules, especially in flavor physics,   
one needs rigorous error estimates of the theoretical results for comparing theoretical predictions 
with the experimental data. 
Systematic errors of the sum-rule results are usually estimated by varying the Borel
parameter and the continuum threshold within some ranges and are believed to be under control. 

The goal of this paper is to study systematic uncertainties of the sum-rule procedure in detail. 
To this end, a quantum-mechanical harmonic-oscillator (HO) potential model is a perfect 
tool (see also \cite{bb}):
in this model both the spectrum of bound states (masses and wave functions) 
and the exact correlator (and hence its OPE to any order) are known precisely. 
Therefore one may apply the sum-rule machinery for extracting parameters of the ground 
state and check the accuracy of the extracted values by comparing with the exact 
known results. In this way the accuracy of the method can be probed.

We show that the knowledge of the correlator in the limited range of the
Borel parameter is not sufficient for a reliable extraction of the
ground-state characteristics from the sum rule, even if the mass of
the ground state is known.  
One should also know the continuum contribution to the correlator with a good accuracy. 

In connection with this observation, we indicate two dangerous points
in a typical sum-rule analysis:

\noindent
(i) A simple modeling of the hadron continuum by a constant effective
continuum threshold leads  
to uncontrolled errors in the extracted hadron parameters. This
occurs even in the case when the true effective continuum threshold 
may be well approximated by a constant, as it happens in the HO model 
considered.

\noindent
(ii) The independence of the extracted ground-state parameter of the 
Borel mass does not guarantee the extraction of its true value. 

\section{The model}
We consider a non-relativistic potential model with the HO potential 
\begin{eqnarray}
\label{1.1}
V(r)=\frac{m\omega^2 \vec r^2}{2}, \qquad r=|\vec r|,   
\end{eqnarray}
and study the polarization operator $\Pi(E)$ defined by 
\begin{eqnarray}
\label{pi}
\Pi(E)=\left(2\pi/m\right)^{3/2}
\langle \vec r_f=0|G(E)|\vec r_i=0\rangle, 
\end{eqnarray}
with $G(E)$ the full Green function of the model, 
\begin{eqnarray}
G(E)=(H-E)^{-1}, \quad H=H_0+V(r), \quad H_0=\vec p^2/2m. 
\end{eqnarray}
The full Green function satisfies the Lippmann-Schwinger operator equation 
\begin{eqnarray}
\label{ls1}
G^{-1}(E)=G_0^{-1}(E)+V, \quad\mbox{ with} \quad G_0(E)=(H_0-E)^{-1},  
\end{eqnarray}
which may be solved perturbatively: 
\begin{eqnarray}
\label{ls2}
G(E)=G_0(E)-G_0(E)VG_0(E)+G_0(E)VG_0(E)VG_0(E)+\cdots.   
\end{eqnarray}
For the polarization operator given by a dispersion representation 
\begin{eqnarray}
\Pi(E)=\int \frac{dz}{z-E} \rho(z), 
\end{eqnarray} 
the Borel transform \cite{svz} has the form 
\begin{eqnarray}
\Pi(\mu)=\int {dz}\exp(-z/\mu) \rho(z).  
\end{eqnarray} 
Therefore the Borel transform corresponds to the evolution operator in the imaginary time $1/\mu$:
\begin{eqnarray}
\Pi(\mu)=\left(2\pi/m\right)^{3/2}
\langle \vec r_f=0|\exp(- H/\mu)|\vec r_i=0\rangle.
\end{eqnarray}
For the HO potential (\ref{1.1}), the exact $\Pi(\mu)$ is known \cite{nsvz}: 
\begin{eqnarray}
\label{piexact}
 \Pi(\mu)=\left(\frac{\omega}{\sinh(\omega/\mu)}\right)^{3/2}. 
\end{eqnarray}
Expanding this expression in inverse powers of $\mu$, we get the OPE series for $\Pi(\mu)$:
\begin{eqnarray}
\label{piope}
\Pi_{\rm OPE}(\mu)\equiv \Pi_{0}(\mu)+\Pi_{1}(\mu)+\Pi_{2}(\mu)+\cdots=
\mu^{3/2}
\left[1-\frac{\omega^2}{4\mu^2}+\frac{19}{480}\frac{\omega^4}{\mu^4}
-\frac{631}{120960}\frac{\omega^6}{\mu^6}
+\cdots \right], 
\end{eqnarray}
and higher coefficients may be obtained from (\ref{piexact}).  
Each term of this expansion may be also calculated from (\ref{pi}) and (\ref{ls2}), 
with $\Pi_0$ corresponding to $G_0$: 
\begin{eqnarray}
\label{pi0}
\Pi_0(\mu)=\int\limits_0^\infty dz\rho_0(z)\exp(-z/\mu),\quad 
\rho_0(z)=\frac{2}{\sqrt{\pi}}\sqrt{z}. 
\end{eqnarray}
The ``phenomenological'' representation for $\Pi(\mu)$ is obtained by using the 
basis of hadron eigenstates of the model, namely
\begin{eqnarray}
\label{piphen1}
\Pi(\mu)=\sum_{n=0}^\infty R_n \exp(-E_n/\mu), 
\end{eqnarray}
with $E_n$ the energy of the $n$-th bound state and $R_n$ 
given by  
\begin{eqnarray}
R_n=(2\pi/m)^{3/2}|\Psi_n(\vec r=0)|^2.
\end{eqnarray}
The quantity $R_n$ determines the square of the leptonic decay constant of the $n$-th bound state. 

For the lowest states one has\footnote{Note that, due to the non-relativistic nature of our HO model, 
the states corresponding to orbital excitations do not contribute to (\ref{pi}) and therefore the 
excited states contributing to (\ref{piexact}) are separated in energy by multiples of $2\omega$ from 
the ground state.}
\begin{eqnarray}
\label{E0}
E_0=\frac{3}{2}\omega,\quad E_1=\frac{7}{2}\omega,\quad \ldots.
\end{eqnarray}
and 
\begin{eqnarray}
\label{r0}
R_0=2\sqrt{2}\omega^{3/2}, \quad R_1=3\sqrt{2}\omega^{3/2},\quad \ldots.
\end{eqnarray}
For later use we isolate the contribution of the ground state and write 
\begin{eqnarray}
\label{piphen}
\Pi(\mu)=R_0 \exp(-E_0/\mu)+\Pi_{\rm cont}(\mu), \qquad 
\Pi_{\rm cont}(\mu)\equiv \int\limits_{z_{\rm cont}}^\infty dz \,\rho_{\rm phen}(z)\exp(-z/\mu), 
\end{eqnarray}
where $\Pi_{\rm cont}$ describes the contribution of the excited states (the model has purely 
discrete
spectrum, but we use the QCD terminology and refer to the excited
states as the  
``continuum''), $z_{\rm cont}$ is the continuum threshold and 
$\rho_{\rm phen}(z)$ is the spectral density corresponding to excited states. 
For the HO potential, the continuum threshold lies at $z_{\rm cont}=\frac{7}{2}\omega$. 

\section{Sum rule}

The sum rule claims the equality of the correlator calculated in the 
``quark'' basis (\ref{piope}) and in the hadron basis (\ref{piphen}):
\begin{eqnarray}
\label{sr}
R_0 e^{-{E_0}/{\mu}}+\int\limits_{z_{\rm cont}}^\infty dz \rho_{\rm phen}(z)e^{-{z}/{\mu}}=
\int\limits_{0}^{\infty}dz \rho_0(z)e^{-{z}/{\mu}} 
+\mu^{3/2}
\left[
-\frac{\omega^2}{4\mu^2}
+\frac{19}{480}\frac{\omega^4}{\mu^4}
-\frac{631}{120960}\frac{\omega^6}{\mu^6}+\cdots\right].
\end{eqnarray} 
Following \cite{nsvz}, we use explicit expressions for the 
power corrections, but for the zero-order free-particle term we use 
its expression in terms of the spectral integral (\ref{pi0}). 
The reason for this will become clear in few lines. 

Let us introduce the effective continuum threshold $z_{\rm eff}(\mu)$, 
different from the physical $\mu$-independent continuum threshold $z_{\rm cont}$, 
by the relation
\begin{eqnarray}
\label{zeff}
\Pi_{\rm cont}(\mu)=
\int\limits_{z_{\rm cont}}^\infty dz \,\rho_{\rm phen}(z)\,\exp(-z/\mu)=
\int\limits_{z_{\rm eff}(\mu)}^\infty dz\, \rho_{0}(z)\,\exp(-z/\mu). 
\end{eqnarray}
Generally speaking, the spectral densities $\rho_{\rm phen}(z)$ 
and $\rho_{0}(z)$ are different functions, so the two sides of (\ref{zeff}) may be equal to
each other only if the effective continuum threshold depends on $\mu$. 
In our model, we can calculate $\Pi_{\rm cont}$ precisely, 
as the difference between the known exact correlator and the known ground-state contribution, 
and therefore we can obtain the function $z_{\rm eff}(\mu)$ by solving (\ref{zeff}) numerically. 
In the general case of a sum-rule analysis, the effective continuum threshold is not known 
precisely and is one of the essential fitting parameters. 

Making use of (\ref{zeff}), we rewrite now the sum rule (\ref{sr}) in the form 
\begin{eqnarray}
\label{sr2}
R_0 \exp({-{E_0}/\mu})=\Pi(\mu,z_{\rm eff}(\mu)),  
\end{eqnarray}
where the cut correlator $\Pi(\mu,z_{\rm eff}(\mu))$ reads
\begin{eqnarray}
\label{cut}
\Pi(\mu,z_{\rm eff}(\mu))\equiv 
\frac{2}{\sqrt{\pi}}\int\limits_{0}^{z_{\rm eff}(\mu) }dz \sqrt{z}\exp(-z/\mu)
+\mu^{3/2}
\left[
-\frac{\omega^2}{4\mu^2}
+\frac{19}{480}\frac{\omega^4}{\mu^4}
-\frac{631}{120960}\frac{\omega^6}{\mu^6}+\cdots\right].  
\end{eqnarray}
As is obvious from (\ref{sr2}), the cut correlator satisfies the equation 
\begin{eqnarray}
\label{e0a}
-\frac{d}{d(1/\mu)}\log \Pi(\mu,z_{\rm eff}(\mu)) =E_0.    
\end{eqnarray}
The cut correlator is the actual quantity which governs the extraction of the ground-state 
parameters. 
It might be useful to notice that the relative weight of power corrections 
in the cut correlator $\Pi(\mu,z_{\rm eff}(\mu))$ has been considerably 
increased compared to the initial $\Pi_{\rm OPE}(\mu)$: whereas 
in $\Pi_{\rm OPE}(\mu)$ power
corrections are suppressed as $1/\mu^2$ compared to the free-particle term, in 
$\Pi(\mu,z_{\rm eff}(\mu))$ they are suppressed only as $1/\sqrt{\mu}$ compared to the
cut free-particle term. In the problem under discussion this makes no difference 
since the power corrections are known precisely. 
In QCD this, however, leads to additional uncertainties since condensates are not 
always known with good accuracy.

The sum rule (\ref{sr2})
allows us to restrict the structure of the effective continuum threshold $z_{\rm eff}(\mu)$. 
Let us expand both sides of (\ref{sr2}) near $\omega/\mu=0$. The l.h.s.\ contains only integer 
powers of $\omega/\mu$, i.e., even powers of $\sqrt{\omega/\mu}$. 
Power corrections on the r.h.s., on the contrary, contain only odd powers of
$\sqrt{\omega/\mu}$. In order that both sides match each other, the effective continuum threshold
should be a power series of the parameter $\sqrt{\omega/\mu}$: 
\begin{eqnarray}
\label{zeff2}
z_{\rm eff}(\mu)=\omega
\left[\bar{z}_0+\bar{z}_1\sqrt{\frac{\omega}{\mu}}+\bar{z}_2\frac{\omega}{\mu}+\cdots\right].       
\end{eqnarray}
Inserting this series in (\ref{sr2}) and expanding the integral on its
r.h.s., we obtain an infinite chain of equations emerging at different orders of
$\sqrt{\omega/\mu}$.
The equations corresponding to the odd powers of $\sqrt{\omega/\mu}$
do not contain the parameters $E_0$ and $R_0$ and 
constrain the odd-number coefficients $\bar{z}_{2i+1}$ which
provide the cancellation of power corrections on the r.h.s.~of
(\ref{sr2}). 
The three lowest-order equations read 
\begin{eqnarray}
&&R_0=\frac{4}{3\sqrt{\pi}}\bar{z}_0^{3/2}\omega^{3/2}=
\int\limits_{0}^{\omega \bar{z}_0}\rho_0(z)dz,  \label{1}         \\
&&\bar{z}_1=\frac{\sqrt{\pi}}{8\sqrt{\bar{z}_0}}, \label{2}\\
&&R_0 E_0=\frac{4}{5\sqrt{\pi}}\bar{z}_0^{5/2}\omega^{5/2}-
\frac{\omega^{5/2}}{2\sqrt{\pi}\sqrt{\bar z_0}}(\bar{z}_1^2+4{\bar{z}_0}\bar{z}_2)=
\int\limits_{0}^{\omega \bar{z}_0}\rho_0(z)z\,dz-
\frac{\omega^{5/2}}{2\sqrt{\pi}\sqrt{\bar{z}_0}}(\bar{z}_1^2+4{\bar{z}_0}\bar{z}_2). \label{3}
\end{eqnarray}
What is essential is that the $i$-th equation contains only the 
variables $\bar{z}_0,\ldots,\bar{z}_i$. 

Setting $E_0=\frac{3}{2}\omega$ and 
$R_0=2\sqrt2\omega^{3/2}$, the equations above yield the following solution for the 
exact effective continuum threshold in the HO model: 
$\bar{z}_0=2.418$, $\bar{z}_1=0.142$, $\bar{z}_2=-0.081$, etc. 

\vspace{.2cm}
The following comments are in order here: 
\begin{itemize}
\item[1.] Equation (\ref{2}) rules out the $\mu$-independent solution $z_{\rm eff}=\mbox{const}$. 

\item[2.] For $E_0$ and $R_0$ within a broad range of values
$0\le R_0 \le R_{\rm upper}$ 
there exists a solution $z_{\rm eff}(\mu,R_0,E_0)$ which 
{\it exactly} solves the sum rule (\ref{sr}). 
Here, the upper boundary $R_{\rm upper}$ is determined from the condition that the 
ground state fully saturates the correlator at $\mu_{\rm min}$, the lower boundary of the 
considered $\mu$-interval: 
\begin{eqnarray}
\label{upper}
R_{\rm upper}\exp({-E_0/\mu_{\rm min}})=\Pi_{\rm OPE}(\mu_{\rm min}). 
\end{eqnarray}
For $\mu_{\rm min}\to 0$, $R_{\rm upper} \to 2\sqrt2\omega^{3/2}$. 

Therefore, in a limited range of $\mu$ the OPE alone cannot say much about the 
ground-state parameters. What really 
matters is the continuum contribution, or, equivalently, $z_{\rm eff}(\mu)$. 
Without constraints on the effective continuum threshold the
results obtained from the OPE are not restrictive.\footnote{The expected sensitivity of the
method should not be overestimated: Imagine, e.g., that we modify the potential as follows: 
$V(r)\to V(r)\exp(-r/r_0)$. Then
the discrete spectrum of states is replaced by a continuous spectrum.
However, for sufficiently large values $r_0\gg 1/\omega$, the power corrections
remain numerically almost unchanged. So the Borel-transformed OPE is not very
sensitive to the dynamics at long distances.}

\item[3.] The approximate extraction of $E_0$ and $R_0$ worked out in a limited range of values 
of $\mu$ becomes possible only by constraining $z_{\rm eff}(\mu)$. If the constraints are 
realistic and turn out to reproduce
with a reasonable accuracy the exact $z_{\rm eff}(\mu)$, then the approximate procedure 
works well. If a good approximation is not found, the approximate procedure fails to
reproduce the true value. Anyway, the accuracy of the extracted value is difficult 
to be kept under control. 
\end{itemize}
The last conclusion is quite different from the results of QCD sum rules presented 
in the literature (see e.g.~the review \cite{ck}). 
In the next section we shall demonstrate that a typical sum-rule analysis contains additional 
explicit or implicit assumptions and criteria for extracting the parameters of the ground state. 
Whereas these assumptions may lead to reasonable central values, the accuracy of the extracted
parameters cannot be controlled. 

\begin{figure}[t]
\begin{center}
\begin{tabular}{cc}
\includegraphics[width=8cm]{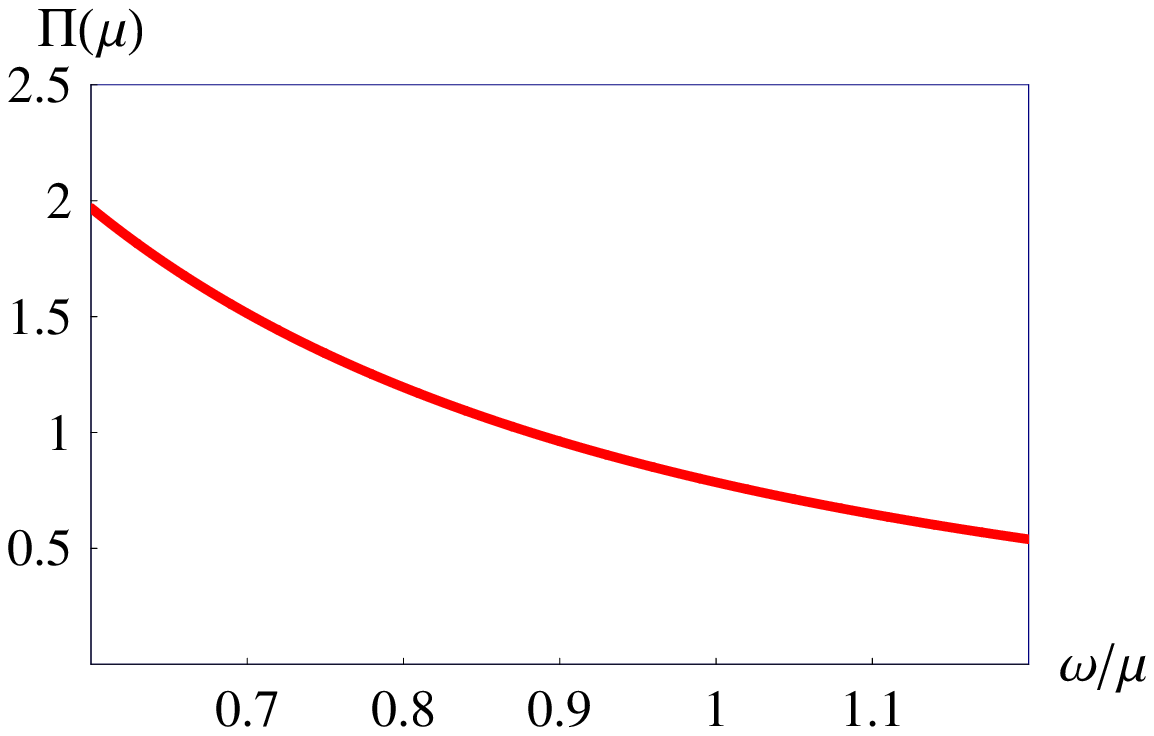}
&
\includegraphics[width=8cm]{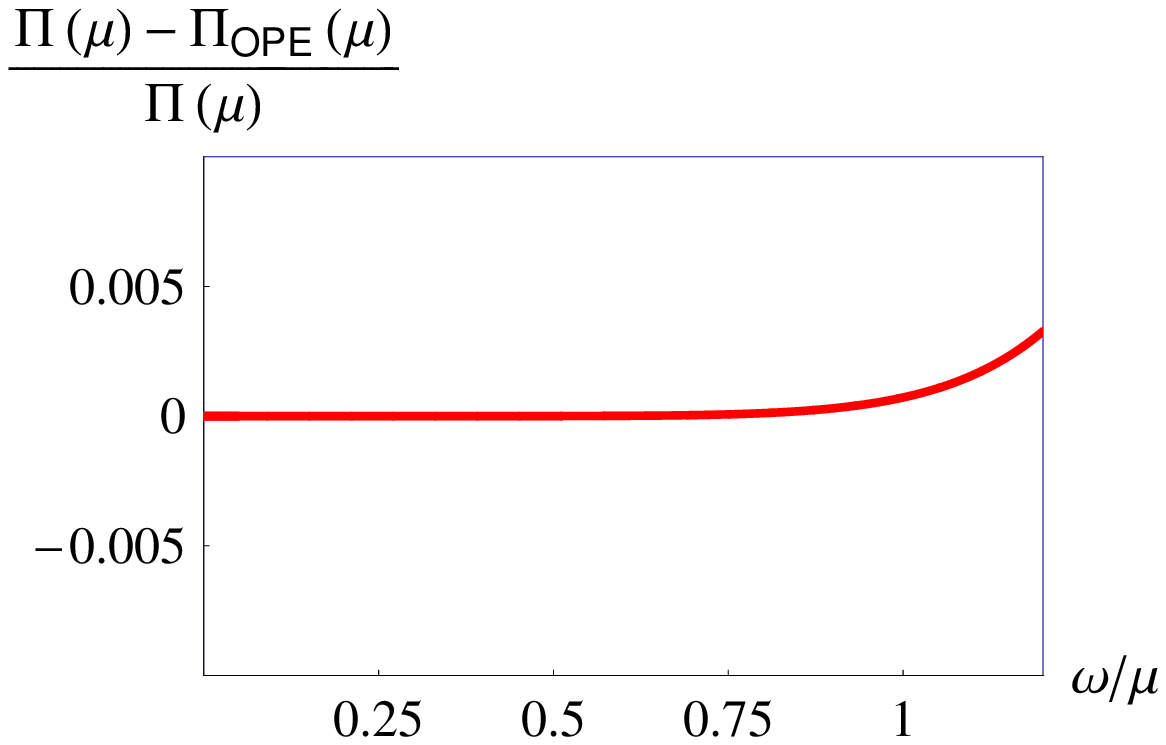}
\end{tabular} 
\caption{\label{Fig:1}
(a) The exact polarization operator $\Pi(\mu)$.  
(b) The accuracy of the OPE: 
the quantity $\left(\Pi(\mu)-\Pi_{\rm OPE}(\mu)\right)/\Pi(\mu)$,
where $\Pi(\mu)$ is the exact 
correlator
and $\Pi_{\rm OPE}$ is the result of the OPE involving the first three power corrections.} 
\end{center}
\end{figure}

\section{Numerical analysis}
In practice, one knows only the first few terms of the OPE, so one 
must stay in a region of $\mu$ bounded from below to guarantee that 
the truncated OPE series reproduces the exact correlator within a controlled accuracy. 
The ``fiducial'' \cite{svz} range of $\mu$ is the range where, on the
one hand, the 
OPE reproduces the exact expression better than some given accuracy, e.g., within 0.5\%, 
and, on the other hand, the ground state is expected to give a sizable contribution to the 
correlator. 
If we include the first three power corrections, $\Pi_1$, $\Pi_2$, and $\Pi_3$, then the fiducial region 
lies at $\omega/\mu<1.2$ (see Fig.~\ref{Fig:1}). Since we know the ground-state parameters, we fix 
$\omega/\mu>0.7$, where the ground state gives more than 60\% of the full correlator. 
So the working range is $0.7<\omega/\mu<1.2$.  

If one knows the continuum contribution with a reasonable
accuracy, one can obviously extract the resonance parameters from the
sum rule (\ref{sr}). We shall be interested, however, in the situation
when the hadron continuum is not known, which is a typical situation in heavy-hadron 
physics and in studying properties of exotic hadrons. 
Can we still extract the ground-state parameters? 

We shall seek the (approximate) solution to the equation 
\begin{eqnarray}
\label{fit}
R \exp({-{E}/\mu})+\int\limits_{z_{\rm eff}(\mu)}^\infty dz \rho_0(z) \exp(-z/\mu)
=\Pi_{\rm OPE}(\mu)  
\end{eqnarray}
in the range $0.7<\omega/\mu<1.2$. 
Hereafter, we denote by $E$ and $R$ the
values of the ground-state parameters as extracted from the sum rule (\ref{fit}). 
The notations $E_0$ and $R_0$ are reserved for the known exact values. 

\subsection{$\mu$-dependent effective continuum threshold}

As already explained, since the continuum contribution to the correlator is positive, 
for any $R$ within the range $0<R<R_{\rm upper}$ there exists a solution
$z_{\rm eff}(\mu,E,R)$, which {\it exactly} solves the sum rule (\ref{fit}). 
Clearly, for different $E$ and $R$ one has a different, specific continuum contribution 
$\Pi_{\rm cont}(\mu,E,R)$. Thus, without measuring $\Pi_{\rm cont}$ or 
imposing constraints on it based on some other considerations  
we cannot extract the ground-state parameters!\footnote{This is a typical situation when one   
studies the existence of exotic states, like tetra- or pentaquarks, with 
QCD sum rules: in this case the relevant continuum is not known, and   
from our point of view, the positive or negative answer to the  
question whether these states exist or not depends mainly on the model used for the continuum.}

In some cases the ground-state energy may be obtained, e.g., from the experiment. However, fixing 
the ground-state energy $E$ equal to its known value $E_0$ does not help much: for any $R$ 
within the range $0 < R < R_{\rm upper}$ one can still find a solution
$z_{\rm eff}(\mu, R)$ which solves the sum rule (\ref{fit}) exactly.

\begin{figure}[b]
\begin{center}
\begin{tabular}{cc}
\includegraphics[width=8.5cm]{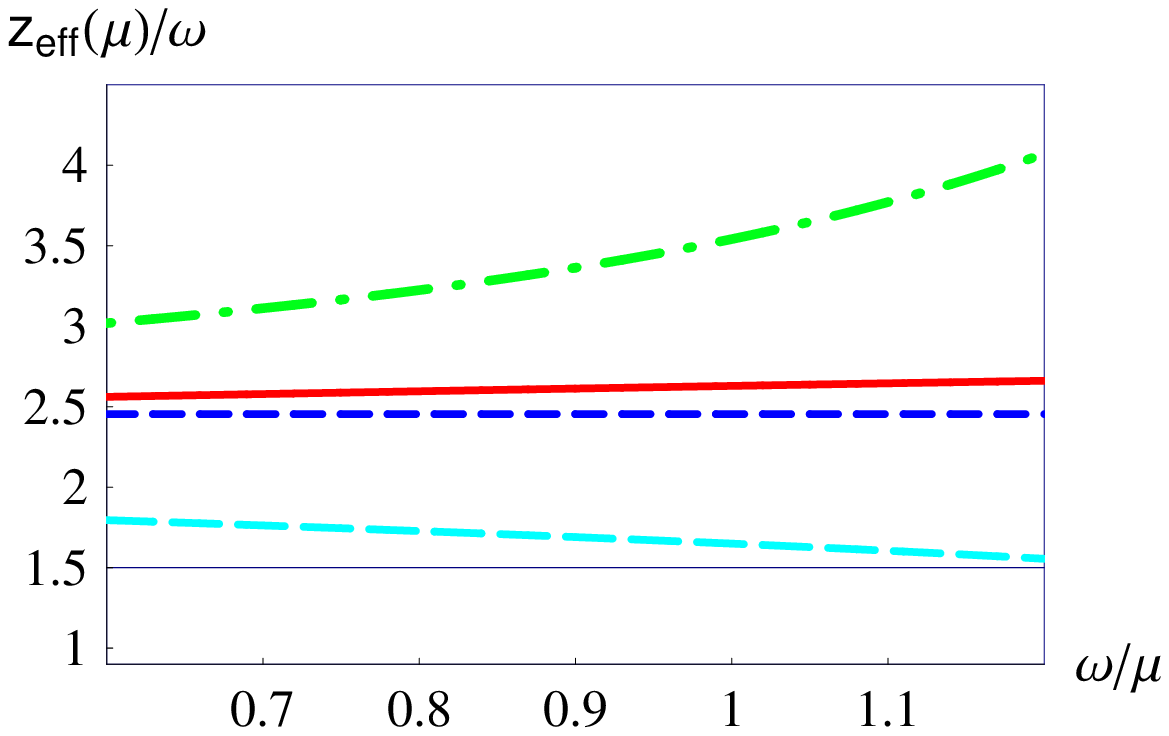}
&
\hspace{.2cm}
\includegraphics[width=8.5cm]{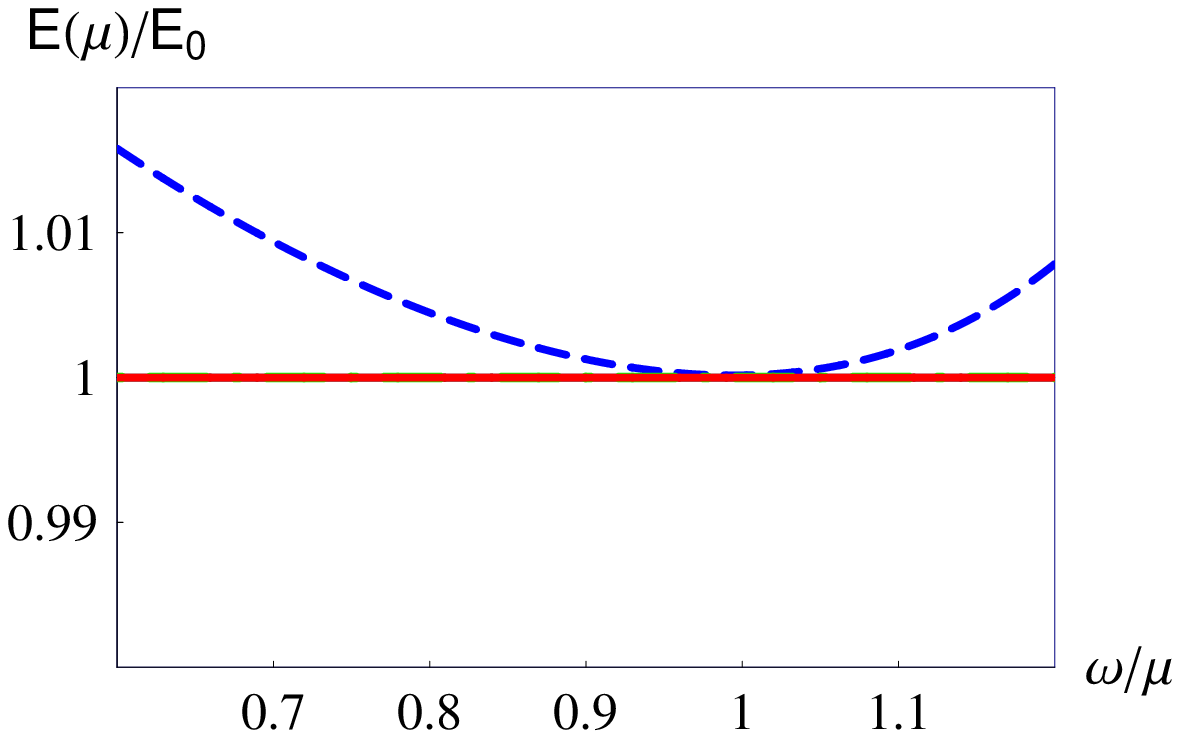}
\\
\includegraphics[width=8.5cm]{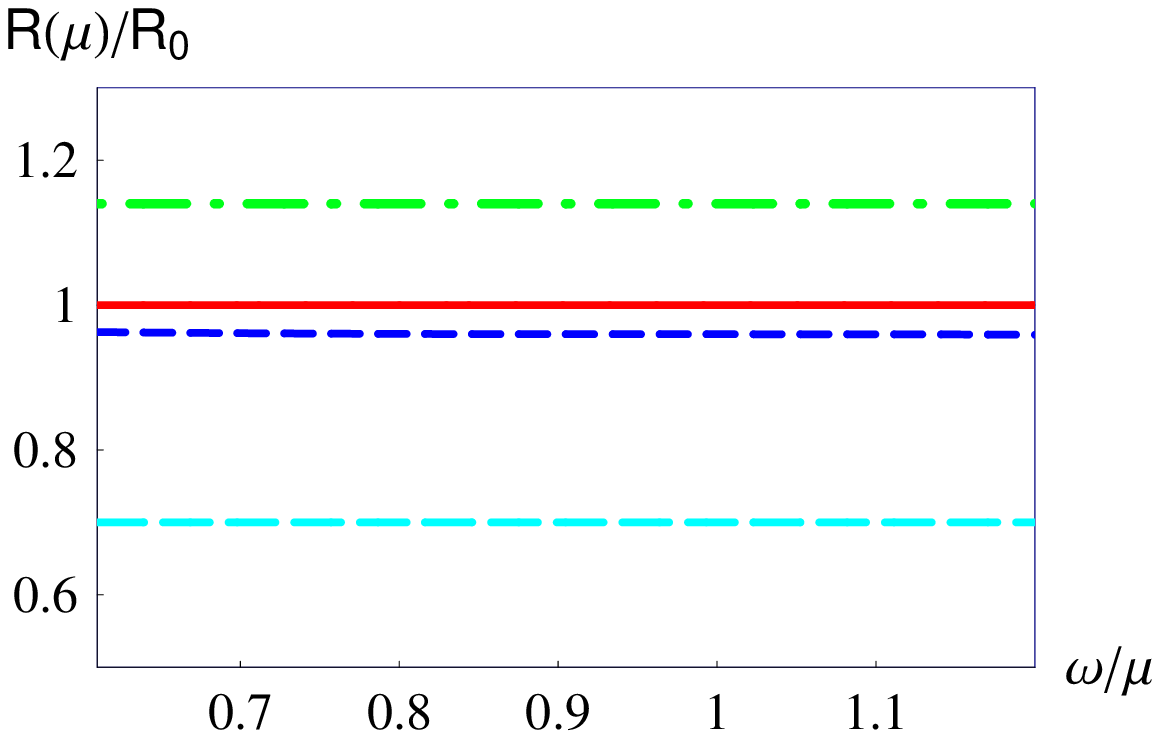}
&
\includegraphics[width=8.5cm]{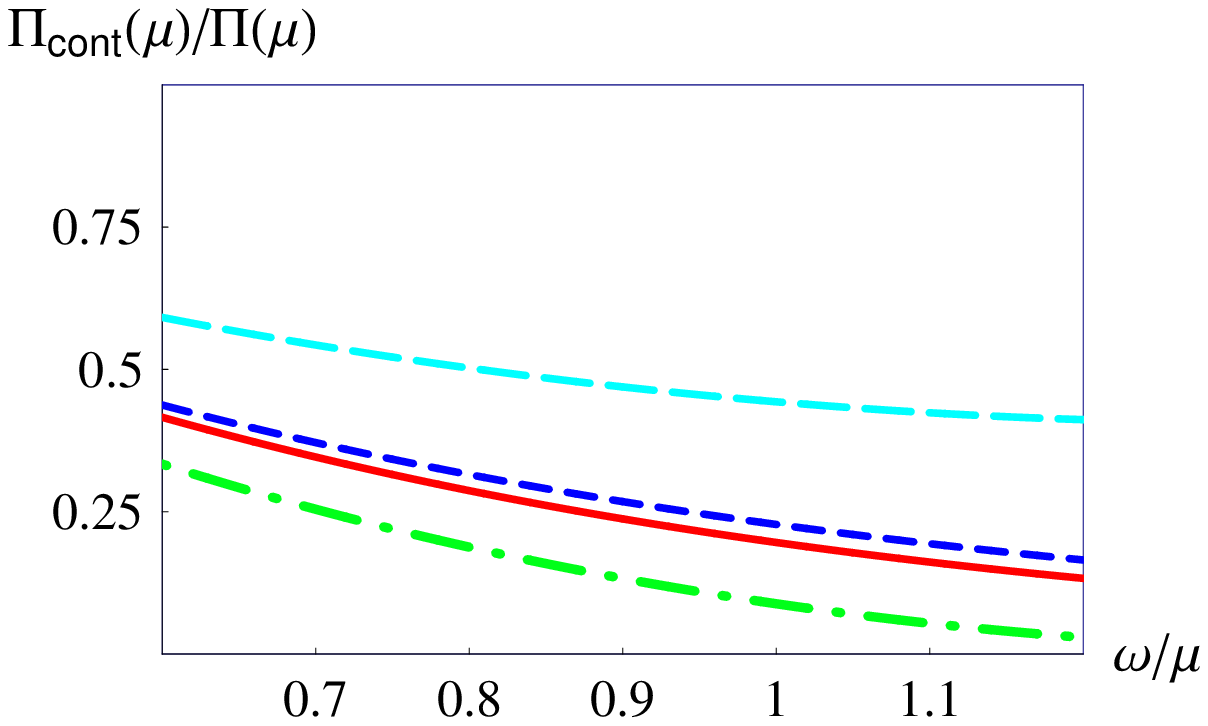}
\end{tabular} 
\caption{\label{Fig:2}
Different choices of the effective continuum threshold $z_{\rm eff}(\mu)$ (a)
and the corresponding $E(\mu)=-d/d\mu \log \Pi(\mu,z_{\rm eff}(\mu))$ (b), 
$R(\mu)$ obtained from the sum rule (\ref{fit}) (c), and $\Pi_{\rm cont}(\mu)$ given by 
Eq.~(\ref{zeff}) (d):
1 [solid (red) line] the exact effective continuum threshold as obtained by a numerical 
solution of (\ref{sr2}), 
2 [long-dashed (blue) line] the effective continuum threshold obtained by 
solving the sum rule (\ref{fit}) for $R=0.7 R_0$ and $E=E_0$,   
3 [dash-dotted (green) line] same as line 2, but for $R=1.15 R_0$ and $E=E_0$,
4 [short-dashed (dark-blue) line] the constant effective continuum threshold $z_c$ 
determined according to Sec.~\ref{constant}. In Plot (b), the lines 1, 2, and 3 
lie one on top of the other and cannot be distinguished.} 
\end{center}
\end{figure}

Let us therefore consider constraints on the effective continuum threshold. 
It is natural to require $z_{\rm eff}(\mu)>E_0$ for all $\mu$. Then the sum rule (\ref{fit}) 
may be solved for any $R$ within the range $0.7<R/R_0<1.15$. The solution 
$z_{\rm eff}(\mu)$ for the boundary values of this interval, and the corresponding 
$E(\mu)=-\frac{d}{d\mu} \log \Pi(\mu,z_{\rm eff}(\mu))$, $R(\mu)$, and 
$\Pi_{\rm cont}$ given by (\ref{zeff}) are shown in Fig.~\ref{Fig:2}. Clearly, 
$\Pi_{\rm cont}$ corresponding to different values of $R$ differ very strongly. 

Fig.~\ref{Fig:2} also presents the exact effective continuum threshold $z_{\rm eff}(\mu)$ 
obtained as a numerical solution of
the sum rule (\ref{fit}) with the known $E_0=\frac{3}{2}\omega$ and $R_0=2\sqrt{2}\omega^{3/2}$, 
and the corresponding $E(\mu)$, $R(\mu)$, and $\Pi_{\rm cont}(\mu)$. 

In the model under discussion, one may expect the exact effective continuum threshold to be 
somewhere between $E_0$ and $E_1$: it is indeed not far from 
$\frac12(E_0+E_1)=\frac{5}{2}\omega$, see Fig.~\ref{Fig:2}(a).
Requiring, e.g., $z_{\rm eff}>\frac12(E_0+E_1)$ gives $0.95<R/R_0<1.15$, 
which is also not too restrictive.

\subsection{Constant effective continuum threshold\label{constant}}

Strictly speaking, the constant effective continuum threshold $z_{\rm eff}(\mu)=z_c={\rm const}$ 
is incompatible with the sum rule, as it can be seen from Eq.~(\ref{2}). 
Nevertheless, this Ansatz may work well, especially in our model: 
as can be seen from Fig.~\ref{Fig:2}(a), the exact $z_{\rm eff}(\mu)$ is almost flat in
the fiducial interval. Therefore, the HO model represents a very favorable situation for
applying the QCD sum-rule machinery.

Now, one needs to impose a criterion for fixing $z_c$. 
One of the widely used ways is the following \cite{jamin}: one calculates 
\begin{eqnarray}
\label{e0b}
-\frac{d}{d(1/\mu)}\log \Pi(\mu,z_c)\equiv E(\mu,z_c).     
\end{eqnarray}
The r.h.s.~depends on $\mu$ due to approximating $z_{\rm eff}(\mu)$ with a constant. 
Then, one determines $\mu_0$ and $z_c$ as the solution to the system
of equations 
\begin{eqnarray}
\label{add}
E(\mu_0,z_c)=E_0,\qquad
\frac{\partial}{\partial\mu}E(\mu,z_c)|_{\mu=\mu_0}=0,  
\end{eqnarray}
yielding the values $z_c=2.454\,\omega$ and $\mu_0/\omega=1$, see
Fig.~\ref{Fig:2}(a,b). 
The central value of the sum-rule estimate $R$ is obtained by setting 
$\mu\to\mu_0$, and $z_{\rm eff}(\mu)\to z_c$ in (\ref{sr2}). For this value of $z_c$ 
one has a 
very good description of $\Pi(\mu)$ (less than 1\% deviation in the full
range $0.7\le \mu/\omega \le 1.2$)   
and the stability of $E(\mu,z_c)$ against $\mu$ is also very satisfactory. 
Finally, in the model under discussion one obtains also a rather good estimate 
$R/R_0=0.96$, with the function 
$R(\mu,z_c)$ being extremely stable in the region $0.7\le \omega/\mu\le
1.2$. 
Note, however, a dangerous point: the description of $\Pi(\mu)$ with
better than 1\% accuracy and the deviation of the 
$E(\mu,z_c)$ from $E_0$ at
the level of only 1\% in the fiducial range leads to a 4\% error
in the extracted value of $R$! 

The crucial conclusion from this observation is the following: 
even when the effective continuum threshold $z_{\rm eff}(\mu)$ is 
almost flat in the fiducial interval
of $\mu$, as in our simple model, one still cannot 
control the accuracy of the extracted value of $R$. As is obvious from Fig.~\ref{Fig:2}(c), 
it would be incorrect to estimate the error, e.g., from the range covered by $R$ 
when varying the Borel parameter $\mu$ within the fiducial interval. 

\subsection{Local-duality limit $\mu\to\infty$}

Let us consider another scheme: a local-duality (LD) sum rule proposed in \cite{radyushkin}.  
This scheme corresponds to the limit $\mu\to\infty$ in (\ref{sr2}) and has several attractive 
features \cite{lm}. 
In the limit $\mu\to\infty$ all power corrections in the OPE vanish and we end up with the simple 
relation (cf.\ Eq.~(\ref{1}))
\begin{eqnarray}
\label{ldsr1}
R_{\rm LD}= 
\int\limits_{0}^{z_{\rm LD}}dz \rho_0(z)=\frac{4}{3\sqrt{\pi}}z_{\rm LD}^{3/2}.  
\end{eqnarray}
Let us consider the average energy calculated with the cut correlator 
\begin{eqnarray}
\label{ldsr2}
E_{\rm LD}=\frac{\displaystyle\int\limits_{0}^{z_{\rm LD}}dz z\rho_0(z)}
{\displaystyle\int\limits_{0}^{z_{\rm LD}}dz \rho_0(z)}
=\frac{3}{5}z_{\rm LD}.    
\end{eqnarray}
It is natural to require $E_{\rm LD}=E_0$. Then 
$z_{\rm LD}=\frac52\omega$ and (\ref{ldsr1}) leads to 
\begin{eqnarray}
\label{ldsr4}
R_{\rm LD}/R_0=\frac{5\sqrt{5}}{6\sqrt{\pi}}\simeq 1.05.    
\end{eqnarray}
As follows from (\ref{1}) and (\ref{3}), the exact values $R_0$ and $E_0$ satisfy the equations   
\begin{eqnarray}
\label{ldsr3}
R_0=\frac{4}{3\sqrt{\pi}}(\omega\bar {z}_0)^{3/2},\qquad
E_0=
\frac{\displaystyle\int\limits_{0}^{\omega \bar{z}_0}dz z\rho_0(z)}
{\displaystyle\int\limits_{0}^{\omega \bar{z}_0}dz \rho_0(z)}
-\frac{3}{8} \omega\frac{\bar{z}_1^2+4{\bar{z}_0}\bar{z}_2}{\bar{z}_0^{2}}.     
\end{eqnarray}
Comparing these equations with (\ref{ldsr1}) and (\ref{ldsr2}), we see that if 
$E_{\rm LD}=E_0$, then $R_{\rm LD}\ne R_0$. This leads to the 
5\% discrepancy in (\ref{ldsr4}). Anyway, the estimate (\ref{ldsr4}) is  
quite good (due to the specific values of the constants $\bar{z}_i$ in the HO model) 
but its accuracy cannot be controlled. 

Closing this section, we note that the issue of the uncertainties
within QCD sum rules 
(see also \cite{LC} for the case of light-cone QCD sum rules) appears 
to be qualitatively similar to what happens in 
other phenomenological approaches, like the constituent quark model,
which have indeed many 
common features with sum rules, as discussed in \cite{ms}. 

\section{Conclusions}
We studied the extraction of the ground-state parameters from the polarization operator 
using various versions of sum rules in the case of the non-relativistic harmonic-oscillator 
potential model. The advantage of such a simple model is that both 
the OPE for the polarization operator and the exact spectrum are known, 
therefore allowing us to compare the results obtained by sum rules with the exact
values and to probe in this way the uncertainties of the method. 

\vspace{.3cm}
Our conclusions are as follows: 
\begin{itemize}
\item
The knowledge of the correlator in a limited range of the Borel
parameter $\mu$ 
is not sufficient for an extraction of the ground-state parameters with a controlled
accuracy, even if the ground-state mass is known precisely:  
Rather different models for the correlator in the form (\ref{fit}) --- a ground state
plus an effective continuum, described by an effective continuum
threshold $z_{\rm eff}(\mu)$ --- lead to the same correlator. 
\item
The procedure of fixing the effective continuum threshold by requiring
that the average mass calculated with the cut correlator (\ref{cut}) should
reproduce the known value of the ground-state mass \cite{jamin,bz} is,
in general, not restrictive: 
a $\mu$-dependent effective continuum threshold $z_{\rm eff}(\mu)$ which solves 
the sum rule (\ref{fit}) leads to the cut correlator
(\ref{cut}) which automatically 
(i) reproduces precisely $E(\mu)=E_0$ for all values of the Borel parameter $\mu$, 
and 
(ii) leads to the $\mu$-independent value of $R$ which, however, may be rather far from 
the true value. 

In the model considered we obtained the following results: 

a. Without constraining $z_{\rm eff}(\mu)$, for any 
value of $R$ within the range $0\le R/R_0\le 1.15$ one can find a function 
$z_{\rm eff}(\mu)$ which exactly solves the sum rule for $0.7\le\omega/\mu\le 1.2$. 

b. Requiring $z_{\rm eff}(\mu)>E_0=\frac32\omega$ for $0.7\le
\omega/\mu \le 1.2$ gives $0.7\le R/R_0\le 1.15$.  

\item 
We studied in detail the standard approximation of the effective
continuum threshold with a constant $z_c$. Within this approximation, 
one can tune the value $z_c$ by requiring that the average 
energy $E(\mu)$ calculated with the cut correlator (\ref{cut}) should reproduce the ground-state
energy $E_0$ in the stability region. In the model under discussion, one obtains in this way a
good estimate $R/R_0=0.96$, with practically $\mu$-independent $R$. 
The unpleasant feature is that the deviation of $R$ from $R_0$ turns out to be much larger
than the variations of $E(\mu)$ and $R$ over the range $0.7\le
\omega/\mu\le 1.2$. And, more importantly, error estimates for $R$ cannot be provided.  
\item
Therefore, we conclude that a sum-rule extraction of the ground-state
parameters without knowing the hadron continuum suffers from uncontrolled systematic 
uncertainties (not to be confused with the uncertainties related to errors in quark masses, 
$\alpha_s$, renormalization point, condensates, etc; the latter errors are usually properly taken 
into account). Unfortunately, a typical sum-rule analysis of heavy-meson observables belongs 
to this class of 
problems: in this case, the hadron continuum is usually not known and is modeled by an effective 
continuum threshold treated as a fit parameter. 
Then, no estimates of systematic errors for the ground-state
parameters obtained with sum rules can be given,  
although the central values may be rather close to the true values. 
Let us also emphasize an important point: as we have demonstrated, the independence of the extracted hadron 
parameters from the Borel mass does not guarantee the extraction of their true values. 
\end{itemize}
We have nevertheless seen that in the model under consideration the sum rules give good 
estimates for the parameter $R_0$. This seems to be due to the
following specific features of the model: 
(i) a large gap between the ground state and the first excitation that contributes to the sum rule;
(ii) an almost constant exact effective continuum threshold in a wide range of $\mu$. 
Whether or not the same good accuracy may be achieved in QCD, where the 
features mentioned above are absent, is not obvious: 
within the standard procedures adopted in QCD sum rules it is practically impossible 
to control the systematic uncertainties of the obtained hadron parameters. 
This shortcoming ---
the impossibility to control the systematic errors --- 
remains the weak feature of the method of sum rules and an
obstacle for using the results from QCD sum rules for precision
physics, such as electroweak physics. 

\vspace{1cm}
\noindent
{\it Acknowledgments.} 
We are grateful to R.~A.~Bertlmann for interesting discussions. 
D.~M.~was supported by the Austrian Science Fund (FWF) under project
P17692. The work was supported in part by RFBR project 07-02-00551a. S.~S.~thanks 
the Institute for High Energy Physics of the
Austrian Academy of Sciences and the Faculty of Physics of the
University of Vienna for warm hospitality.

\end{document}